\documentclass{llncs}
\pdfoutput=1

\usepackage{booktabs}
\usepackage[sort,sectionbib,round]{natbib}

\usepackage{microtype}
\usepackage{url}

\usepackage{graphicx}

\title{Reliability of Computational Experiments on Virtualised Hardware}

\author{Ian P. Gent and Lars Kotthof\/f\\
\texttt{\{ipg,larsko\}@cs.st-andrews.ac.uk}}
\institute{University of St Andrews}

\begin{document}

\maketitle

\begin{abstract}
We present preliminary results of an investigation into the suitability of
virtualised hardware -- in particular clouds -- for running computational
experiments. Our main concern was that the reported CPU time would not be
reliable and reproducible. The results demonstrate that while this is true in
cases where many virtual machines are running on the same physical hardware,
there is no inherent variation introduced by using virtualised hardware compared
to non-virtualised hardware.
\end{abstract}

\section{Introduction}

Running computational experiments is a task that requires a lot of resources.
Especially recent research in Artificial Intelligence is concerned with the
behaviour of a large number of problem-solving systems and algorithms on a
large number of problems \citep{satzilla,ensemble}. The purpose of these
large-scale experiments is to build statistical models of the behaviour of
certain systems and algorithms on certain problems to be able to predict the
most efficient system for solving new problem instances.

The obvious problem is that a lot of computing resources are required to be able
to run this kind of experiments. Provisioning a large number of machines is not
only expensive, but also likely to waste resources when the machines are not
being used. Especially smaller universities and research institutions are often
unable to provide large-scale computing infrastructure and have to rely on
support from other institutions.

The advent of publicly available cloud computing infrastructure has provided a
possible solution to this problem. Instead of provisioning a large number of
computers themselves, researchers can use computational resources provided by
companies and only pay for what they are actually using. Nowadays commercial
clouds are big enough to easily handle the demand running large-scale
computational experiments generates.

This raises an important question however. How reliable and reproducible are the
results of experiments run in the cloud? Are the CPU times reported more
variable than on non-virtualised hardware?

While the focus of our evaluation is on computational experiments, we believe
that the results are of interest in general. If a company is planning the
provisioning of virtual resources, the implicit assumption is that the
performance of the planned resources can be predicted based on the performance
of the already provisioned resources. If these predictions are unreliable, too
few resources could be provisioned, leading to a degradation of performance,
or too many, leading to waste.

\section{Related work}

There has been relatively little research into the repeatability of experiments
on virtualised hardware. \cite{exploring} report large fluctuations of
high-performance computing workloads on cloud infrastructure. \cite{cloudperf}
evaluate the performance of the Amazon cloud with regards to its general
suitability for scientific use.  The handbook of cloud
computing \citep{cloudhand} explores the issue in some of its chapters.

An experimental evaluation by \cite{cloudruntime} again showed that there is
large variability in performance and care must be taken when running scientific
experiments. They provide an in-depth analysis of the various factors that
affect performance, but only distinguish between two different virtual machine
types provided by the Amazon cloud.

Our approach is more systematic and directly compares the variability of
performance on virtualised and non-virtualised hardware with a real scientific
workload. Our application is lifted straight from Artificial Intelligence
research.

\section{Problem statement}

We are concerned with two major problems when running experiments. First, we
want the results to be \textbf{reliable} in the sense that they faithfully
represent the true performance of an algorithm or a system. Second, we want them
to be \textbf{reproducible} in the sense that anybody can run the experiments
again and achieve the same results we did.

We can assess the reliability of an experiment by running it several times and
judging whether the results are the same within some margin of experimental
error.  Reproducibility is related to this notion, but more concerned with being
able to reproduce the results in a different environment or at a different time.
The two concepts are closely related however -- if we cannot reproduce the
results of an experiment it is also unreliable and if the results are unreliable
there is no point in trying to reproduce them.

Running experiments on virtualised hardware gives an advantage in terms of
reproducibility because the environment that an experiment was run in can be
packaged as a virtual machine. This not only removes possible variability in the
results due to different software versions, but also enables to reproduce
experiments with unmaintained systems that cannot be built and would not run on
contemporary operating systems.

The questions we investigate in this paper however are as follows.
\begin{itemize}
\item Is there inherently more variation in terms of CPU time on virtualised
    hardware than on non-virtualised hardware?
\item Is the performance of virtualised hardware consistent and are we able to
    combine several virtual machines into a cluster and still get consistent
    results?
\item Are there differences between different clouds that use different
    controller software?
\end{itemize}


\section{Experimental evaluation}

To evaluate the reliability of experimental results, we used the Minion
constraint solver \citep{minion}. We ran it on the following three problems.
\begin{itemize}
\item An $n$-queens instance that takes a couple of seconds to solve (place $n$
    queens on an $n\times n$ chessboard such that no queen is attacking another
    queen).
\item A Balanced Incomplete Block Design (BIBD) problem that takes about a
    minute to solve, CSPLib \citep{csplib} problem 028.
\item A Golomb Ruler problem that takes several hours to solve, CSPLib problem
    006.
\end{itemize}

There is a large variation of CPU time across the different problems. This
enables us to isolate short-term effects (such as virtualisation of CPUs) from
long-term effects (such as other jobs the operating system runs overnight).

We ran the experiments in three different settings --
\begin{itemize}
\item on three 8-core machines with non-virtualised hardware,
\item on the Eucalyptus-based private StACC
    cloud\footnote{\url{http://www.cs.st-andrews.ac.uk/stacc}} and
\item on the public Amazon cloud.
\end{itemize}

For the Amazon cloud, we investigated the different virtual machine types
\texttt{m1.large}, \texttt{m1.xlarge}, \texttt{c1.xlarge} and
\texttt{m2.4xlarge}\footnote{\url{http://aws.amazon.com/ec2/instance-types/}}.
In each case, we provided 16 cores to run the experiments, i.e.\ 8 different
virtual machines for \texttt{m1.large} and 2 different virtual machines for
\texttt{m2.4xlarge}. In the StACC cloud, we used 5 virtual machine instances
with 2 cores each.

Using several virtual machines introduces an additional source of variation, but
at this stage of the evaluation we are interested in the reliability of
experimental results that require a large amount of resources and therefore
several machines.

The experiments on non-virtualised hardware establish the baseline of
reliability we can expect. We can then compare the reliability on virtualised
hardware to see if it is significantly worse. Each problem was solved 100 times.
We used the coefficient of variation (standard deviation divided by mean) of the
CPU time required to solve a problem across the 100 runs as a measure of the
reliability of the results.

\section{Results and analysis}

The results for all problems and experimental settings are summarised in
Table~\ref{res}. We were surprised to find that the coefficient of variation of
the reported CPU time on the largest virtual machine type in the Amazon cloud
was lower than what we achieved on non-virtualised hardware. This demonstrates
that running on virtualised hardware does not introduce additional variability
per se.

\begin{table}
\centering
\begin{tabular}{llll}
experimental setting & $n$-queens & BIBD & Golomb Ruler\\\midrule
non-virtualised & 0.016 & 0.018 & 0.005\\
StACC & 0.013 & 0.022 & 0.009\\
Amazon \texttt{m1.large} & 0.333 & 0.13 & 0.183\\
Amazon \texttt{m1.xlarge} & 0.264 & 0.235 & 0.271\\
Amazon \texttt{c1.xlarge} & 0.055 & 0.028 & 0.042\\
Amazon \texttt{m2.4xlarge} & \textbf{0.008} & \textbf{0.008} & \textbf{0.003}\\
\end{tabular}
\caption{Coefficient of variation for all experiments. The lowest figures for
each problem are in \textbf{bold}.}
\label{res}
\end{table}

We furthermore observed the general trend of the coefficient of variation
decreasing as the experiment takes longer to run. This does not seem to be true
on virtual machine types that have a large coefficient of variation though.
Overall, the differences between the different experimental settings are two
orders of magnitude. This is an indication that evaluations like this one are
necessary and we cannot assume that the performance of any given virtual machine
will be consistent and reliable.

\begin{figure*}
\includegraphics[width=\textwidth]{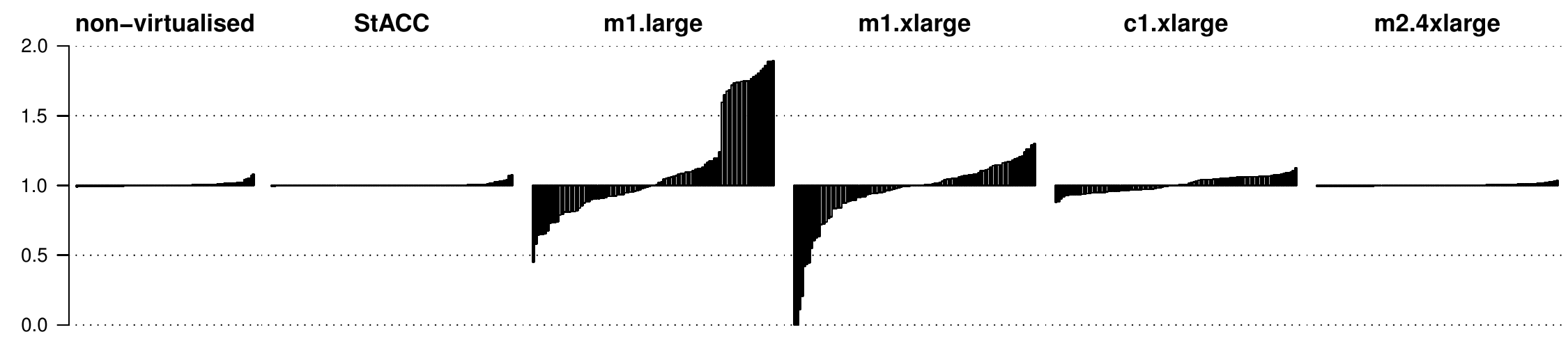}
\caption{Relative deviation from the median CPU time for the $n$-queens
problem for each run. 1 is the median value, 2 means that the run took twice as
long as the median and 0 means that it took no time.}
\label{var-queens}
\end{figure*}

\begin{figure*}
\includegraphics[width=\textwidth]{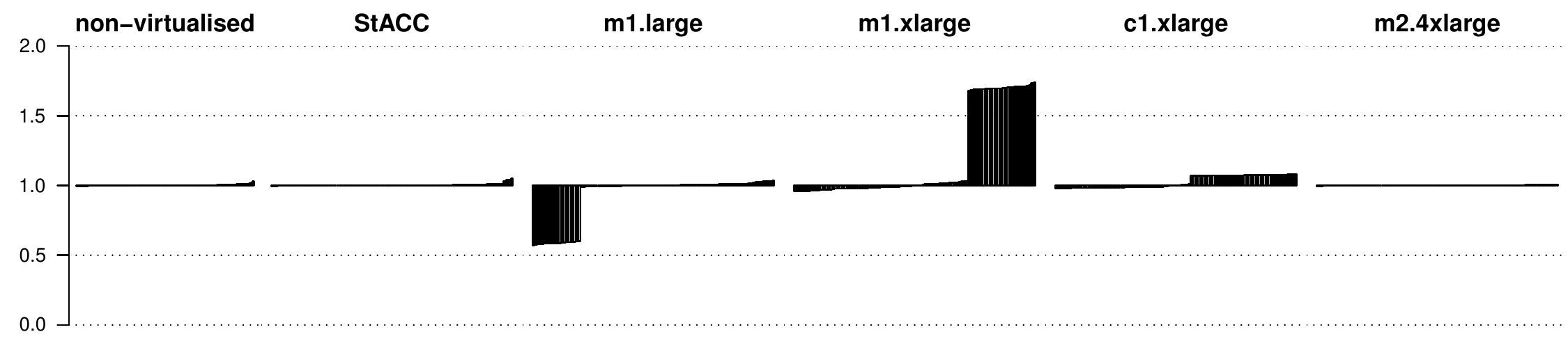}
\caption{Relative deviation from the median CPU time for the Golomb Ruler
problem for each run.}
\label{var-golomb}
\end{figure*}

The variation for each individual run is depicted in Figure~\ref{var-queens} for
the $n$-queens problem and Figure~\ref{var-golomb} for the Golomb Ruler problem.
The distribution for the $n$-queens problem, which takes only a few seconds to
solve, is more or less uniform. For the Golomb Ruler, which takes several hours
to solve, however, there are distinct plateaus. We believe that these are caused
by the different virtual machines we used. That is, two of the eight virtual
machines used of type \texttt{m1.large} were significantly slower than the rest.
Such a difference is still visible for type \texttt{c1.xlarge}, where two
different virtual machine instances were used. There is no noticeable difference
between the two \texttt{m2.4xlarge} instances however.

The coefficient of variation of the Eucalyptus-based StACC cloud is very similar
to the one on non-virtualised hardware and not significantly better or worse
than that of the Amazon cloud.

\section{Conclusions and future work}

We have presented the results of a preliminary evaluation of the variation of
CPU time on virtualised vs.\ non-virtualised hardware. We can draw the following
conclusions.
\begin{itemize}
\item The differences in variation across different types of virtual machines
    and non-virtualised hardware can be several orders of magnitude.
\item Virtualised hardware does not introduce additional variation compared to
    non-virtualised hardware per se. This does not hold true for all types of
    virtual machines however.
\item Performance varies across different instances of the same virtual machine
    type, but the variation decreases for larger virtual machine types.
\item There does not appear to be a significant difference between different
    cloud systems (StACC Eucalyptus cloud and Amazon cloud).
\end{itemize}

The variation of CPU times on the largest virtual machine type on the Amazon
cloud (\texttt{m2.4xlarge}) is at least as good as on non-virtualised hardware.
In terms of reliability of results, it is therefore a feasible alternative to
physical hardware to run experiments on. The high price of this instance type
however eliminates some of the benefits of not having to provision hardware and
paying only for what is actually used.

In the future, we are planning on investigating the variation between different
virtual machines of the same type further; especially across different data
centres. We are also planning on investigating the repeatability of
experimental results over time. The evaluation of the financial feasibility is
another important subject for future research.

\section*{Acknowledgments}

This research was supported by a research grant from Amazon Web Services. Lars
Kotthof\/f is supported by a SICSA studentship.

\bibliographystyle{plainnat}
\renewcommand\bibname{References}
\bibliography{experiments}

\end{document}